%% file: 0_paper.tex
\documentclass{article}

\usepackage{arxiv}

\usepackage[utf8]{inputenc} 
\usepackage[T1]{fontenc}    
\usepackage{hyperref}       
\usepackage{url}            
\usepackage{booktabs}       
\usepackage{amsfonts}       
\usepackage{nicefrac}       
\usepackage{microtype}      
\usepackage{lipsum}
\usepackage{graphicx}
\usepackage{multirow}
\usepackage{authblk}

\usepackage{mathtools}

\DeclarePairedDelimiter\floor{\lfloor}{\rfloor}

\title{Automated Cardiothoracic Ratio Calculation and Cardiomegaly Detection using Deep Learning Approach}

\author[1]{Isarun Chamveha}
\author[1,2]{Treethep Promwiset}
\author[3]{Trongtum Tongdee}
\author[3]{Pairash Saiviroonporn}
\author[2]{Warasinee Chaisangmongkon}
\affil[1]{Perceptra Co., Ltd., Bangkok, Thailand}
\affil[ ]{\texttt{isarun@perceptra.tech}}
\affil[2]{Institute of Field Robotics, King Mongkut's University of Technology Thonburi, Bangkok, Thailand}
\affil[ ]{\texttt{\{tretap.fibo, warasinee.cha\}@mail.kmutt.ac.th}}
\affil[3]{Radiology Department, Faculty of Medicine Siriraj Hospital, Mahidol University, Bangkok, Thailand}
\affil[ ]{\texttt{\{trongtum, pairash.sai\}@gmail.com}}

\begin{document}
\maketitle

\begin{abstract}
We propose an algorithm for calculating the cardiothoracic ratio (CTR) from chest X-ray films. Our approach applies a deep learning model based on U-Net with VGG16 encoder \cite{balakrishna2018automatic} to extract lung and heart masks from chest X-ray images and calculate CTR from the extents of obtained masks.
Human radiologists evaluated our CTR measurements, and $76.5\%$ were accepted to be included in medical reports without any need for adjustment. This result translates to a large amount of time and labor saved for radiologists using our automated tools.
\end{abstract}

\input{1_introduction.tex}
\input{2_litreview.tex}
\input{3_methods.tex}

\input{4_results.tex}

\input{5_conclusion.tex}
\bibliography{references}{}
\bibliographystyle{unsrt}

\end{document}

%% file: 1_introduction.tex
\section{Introduction}

Chest X-ray, or CXR, is widely used in diagnosing abnormal conditions in the chest and nearby structure. Radiologists routinely perform cardiothoracic ratio (CTR) measurement on antero-posterior chest radiographs to diagnose Cardiomegaly, a condition that is strongly correlated with both congenital and congestive heart diseases. Although most Picture Archiving and Communication Systems (PACS) include drawing tools to aid the assessment of CTR, the process is still often labor intensive and time consuming. Manual labeling of organ boundaries and calculation of CTR is prone to error and can lead to faulty interpretations.

Recent advances in machine learning have introduced a wide variety of computer vision methods that can be used to aid this process. Deep learning is a growing trend in medical image analysis, and convolutional neural networks have been shown to yield competitive performances in automated detection of diseases in X-ray images. 


In this work, we explore an automatic approach to calculate CTR from chest X-ray film utilizing deep learning models. Using an image segmentation network based on U-Net with VGG16 encoder \cite{balakrishna2018automatic}, we extract lung and heart regions from the CXR film. The heart and lung diameters are then calculated from the segmented regions.

%% file: 2_litreview.tex
\section{Background}

\subsection{Cardiothoracic Ratio}

\begin{figure}
    \centering
    \includegraphics[width=0.4\linewidth]{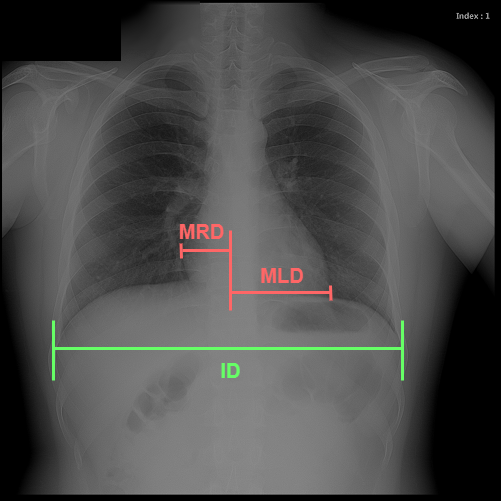}
    \caption{Figure depicting MRD, MLD, and ID measurements}
    \label{fig:ctr_details}
\end{figure}

Chest radiography is the most common tool for detecting cardiomegaly due to the low cost and high availability of devices to obtain chest X-ray images. From chest X-ray images, radiologists employ CTR as one of the most important indicators of cardiomegaly due to the simplicity of the calculation.
CTR of a chest X-ray image is calculated as cardiac diameter (the diameter of the heart) divided by the thoracic diameter (the diameter of the chest).
Specifically, CTR can be calculated from three measurements, $MRD$, the midline-to-right heart diameter, $MLD$, the midline-to-left heart diameter, and $ID$, the internal diameter of chest \cite{mensah2015ctrintro,dimopoulos2013ctrintro}, as
\begin{equation}
CTR = (MRD + MLD) / ID,
\label{eqn:ctr}
\end{equation}
where $MRD$ and $MLD$ is measured from the greatest perpendicular diameter from midline to right and left heart border, respectively. Figure \ref{fig:ctr_details} visualizes the details of CTR calculation. A CTR value of $0.5$ is generally considered to indicate the upper limit of normal.

Most Picture Archiving and Communication Systems (PACS) used by radiologists include tools akin to rulers to ease the process for obtaining these measurements. However, this manual process is labor and time consuming, and can be error-prone when radiologists need to assess hundreds of chest X-ray films a day.

\subsection{Automated CTR Calculation}


There have been several attempts on automatic measurements of cardiothoracic ratio. These approaches involve the calculation of lung and heart regions in the image and use extents of the masks to calculate CTR in a similar practice radiologists use to assess CTR from chest X-ray films.

For the automated segmentation of lung and heart, traditional image processing methods can achieve great results. In an early work, a reference image repository was searched for the nearest neighbor of the patient's sample X-ray image, and a SIFT flow algorithm was used to align and transform lung boundary from the nearest neighbor image to the sample \cite{dallal2017imgproc}.


Ebenezer and Rao \cite{ebenezer2017imgproc} applied a Euler number-based approach to find the best threshold that separates the two lungs from the background. After removing background regions from four corners of the image and applying dilation and erosion for image smoothing, thoracic diameter, $ID$, is then calculated by scanning for leftmost and rightmost points on the lung mask. $MRD$ and $MLD$ are then calculated from the widest point between two lung masks.

Candemir et al. \cite{candemir2016svm} registers the input chest image with the most similar image in the model dataset. The similarity is measured by calculating Bhattacharyya distance of the X-ray intensity histograms. A correspondence map is then calculated using a SIFT flow algorithm to compute a transformation matrix, which is applied to the model mask to transform it into the input image space. CTR value is then calculated from the boundaries of lung and heart masks.

Recent works show excellent results by applying a deep learning approach called U-Net to extract lung and heart boundaries \cite{que2018cardioxnet}. This approach obtained $93.75\%$ accuracy on cardiomegaly detection task, on the dataset of $103$ images from NIH Chest X-ray Dataset \cite{wang2017nihdataset}.
Li et al. \cite{li2019automatic} used U-Net to segment heart and lung masks. They applied a Conditional Random Field to the masks to smooth region boundaries and calculate CTR by measuring the cardio and thoracic diameters from the lung and heart masks. They performed the test on $5,000$ postero-anterior (PA) chest X-ray images from the Radiology Imaging Center in their hospital and obtained $95.3\%$ accuracy on cardiomegaly detection.
Given the success of image segmentation on CTR calculation, we further explore image segmentation approaches using deep learning.


\subsection{Image Segmentation with Deep Learning}
Image segmentation has been one of the most active and sought-after fields of research due to its various applications such as scene understanding, content-based image retrieval, or medical imaging. In the early years, computer vision-based approaches were established for various image segmentation tasks. Although these techniques are accurate on their target tasks, many of them are not easy to train and adapt to new tasks. We refer readers to a comprehensive survey by Zaitoun \cite{zaitoun2015imsegsurvey} for further information on early approaches to image segmentation.

A major breakthrough in image segmentation was made by the introduction of the deep learning approach.
U-Net \cite{olaf2015unet} uses a deep learning model to tackle pixel-wise segmentation tasks accurately with great speed on various segmentation tasks. U-Net employs an end-to-end encoder-decoder network that contains the encoder that performs feature extraction from image input and the decoder that processes the features into output mask. U-Net also concatenates high-resolution features from encoder parts to the decoder network to improve model localization.
Segnet \cite{badrinarayanan2017segnet} is an efficient multi-class segmentation deep learning architecture for scene understanding. Segnet improves computation time and memory usage with some trade-offs in segmentation accuracy. Segnet structure consists of encoder and decoder parts. The encoder includes a sequence of convolutional layers with batch normalization and relu activation function, and the decoder part mirrors the structure of the encoder.

Iglovikov \cite{iglovikov2018ternausnet} and Balakrishna \cite{balakrishna2018automatic} further improve the encoder part with VGG11 and VGG16 networks, respectively.


%% file: 3_methods.tex
\section{Methodology}

\subsection{Dataset}

To construct our heart mask and lung mask dataset, we obtained $245$ images annotated with heart and lung masks from JSRT dataset \cite{shiraishi2000jsrt}.
We obtained $138$ additional lung masks from Montgomery County X-ray dataset \cite{jaeger2014montgomery} and manually segmented additional heart masks by randomly selecting $25$ images with cardiomegaly label and $25$ images with cardiomegaly-negative labels from both the NIH Chest X-ray dataset \cite{wang2017nihdataset} and the CheXpert dataset \cite{irvin2019chexpert}. After filtering out images with poor quality, our manual segmentation of heart masks yielded $86$ additional heart masks.
In total, our dataset contains $383$ lung masks and $331$ heart masks.

After applying histogram equalization to normalize the images (described below), we use image augmentation to increase the number of training data and improve training performance.
Samples were augmented with a random $-8$ to $+8$ degree rotation and randomly apply Gaussian noise and Gaussian blur. We also apply the horizontal flip to lung samples. We obtained a total of $2,002$ heart samples and $1,238$ lung samples and applied a 90-10 train-validation split to divide the samples.

\subsection{Data Preprocessing}
Chest X-ray images in the dataset are taken with different machines across multiple hospitals, therefore, their image intensity varies and needs to be normalized before feeding into a deep learning model. We apply Histogram equalization to normalize images. Specifically, we modify the value of each pixel with the following function:
\begin{equation}
    g_{i,j} = \floor*{(L-1)\sum_{n=0}^{f_{i,j}} p_{n}} ,
\end{equation}
where $g_{i,j}$ is the output intensity at location $(i,j)$, $f_{i,j}$ is the intensity of the original image at location $(i,j)$, $p_n$ is the number of pixels with intensity n divided by total number of pixels, and $L$ is the maximum intensity of the image.

\subsection{Segmentation Model Training}



To select the best approach to obtain heart and lung segment, we compared the accuracy of three image segmentation models: the Segnet \cite{badrinarayanan2017segnet}, the U-Net \cite{olaf2015unet} and the U-Net with VGG16 encoder \cite{balakrishna2018automatic}.





Segnet is a multi-class pixel-wise segmentation network. Segnet structure consists of encoder and decoder sections, and each includes a sequence of convolutional layers with batch normalization and relu activation function. The decoder part mirrors the structure of the encoder. Images are consecutively reduced in size by non-overlapping maxpooling layers through the encoder and gradually restored up to the original size by upsampling layers in the decoder, using maxpooling indices stored from the encoder part.

\begin{figure*}[htbp]
\centering
\includegraphics[width=0.9\linewidth]{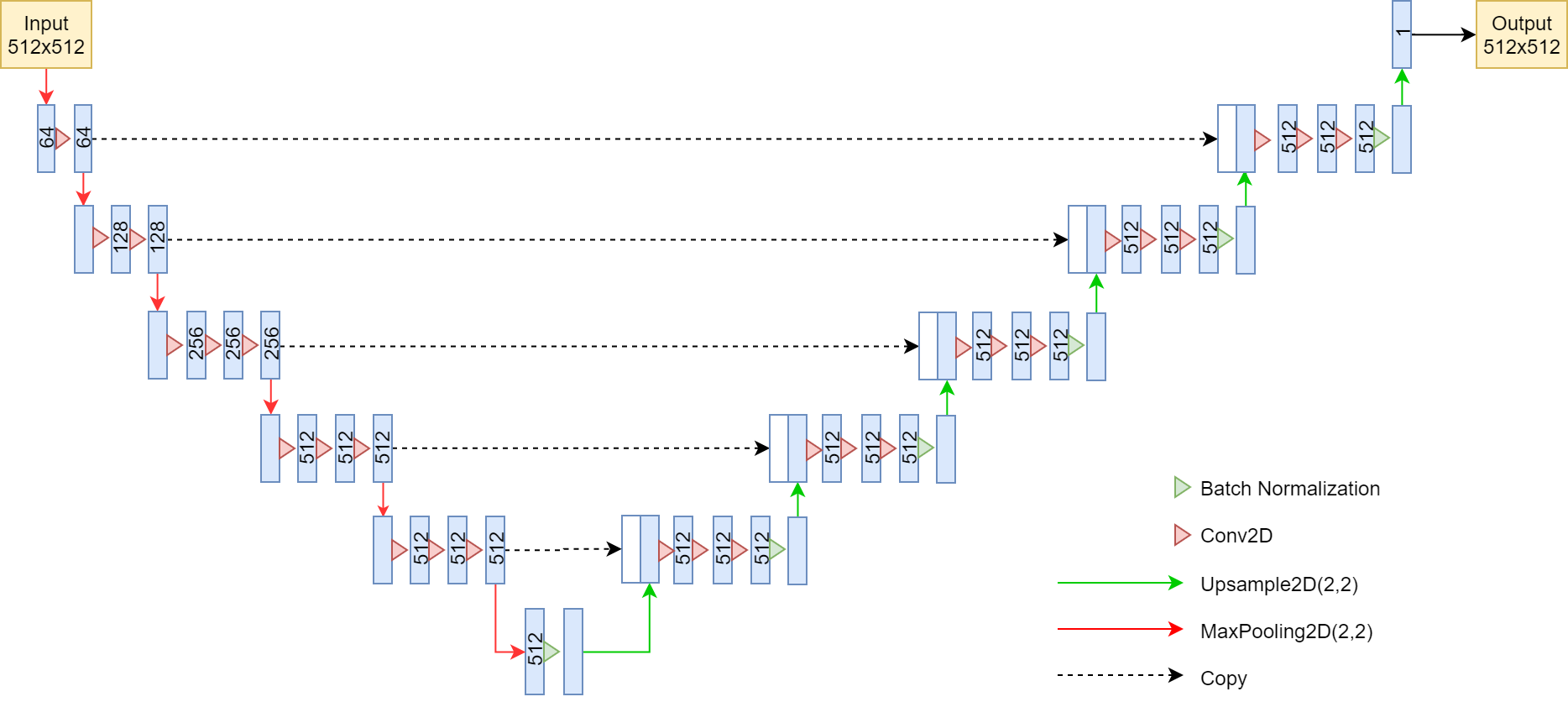}
\caption{Structure of U-Net with VGG16 Encoder}
\label{fig:vgg16_u_net}
\end{figure*}

U-Net is a popular biomedical image segmentation network. Similarly to Segnet, U-Net consists of an encoder, using VGG11 architecture \cite{simonyan2014vgg16}, and a decoder with upsampling layers. The output of each encoder layer is fed into the corresponding decoder layer by concatenating with the output from the previous decoder layer. We modified U-Net by using VGG16 architecture instead of VGG11. Figure \ref{fig:vgg16_u_net} illustrates our model architecture. The input images to our model are 512 by 512 pixels in size.

For loss function, we use a combination of soft dice and binary cross-entropy with logits loss.

\subsubsection{Soft Dice Loss}
Soft dice loss measures the overlap between two mask samples, ranging from 0, where two masks overlap completely, and 1, where there are no overlapping parts between the two masks. Soft dice loss function, $L_{Dice}$, is written as follows:

\begin{equation}
    L_{Dice} = 1-\frac{2\sum_{n}y_{n}p_{n}}{\sum_{n}y_{n}^{2} + \sum_{n}p_{n}^{2}} ,
\end{equation}
where $y_{n}$ is the value of target mask at pixel $n$, and $p_{n}$ is the value of prediction mask at pixel $n$.

\subsubsection{Binary Cross-Entropy With Logits Loss}
Binary cross-entropy with logits loss combines a sigmoid layer and the binary cross-entropy loss to measure the difference between prediction mask and ground-truth mask. We first measure pixel-wise loss by:
\begin{equation}
    l_{n} = y_{n}\cdot\log(p_{n}) + (1-y_{n})\cdot\log(1-p_{n}) ,
\end{equation}
where $l_{n}$ is the loss value of pixel $n$.

The loss function is then calculated as a mean value of all pixels
\begin{equation}
    L_{BCE} = \frac{\sum_{n=1}^{n=N}l_{n}}{N} .
\end{equation}
Our loss function is a sum of the soft Dice loss and the binary cross-entropy with logits loss as
\begin{equation}
    L=L_{Dice}+L_{BCE} .
\end{equation}
Heart mask and lung mask models are trained separately. We trained each model using Adam (Adaptive Moment Estimation) optimizer with a batch size of $8$ for $75$ epochs and an initial learning rate of $0.0001$. Training algorithm is implemented on Nvidia Tesla V100 GPU with 32GB memory.


\subsection{Postprocessing}

\begin{figure}[htbp]
\centering
\includegraphics[width=0.4\linewidth]{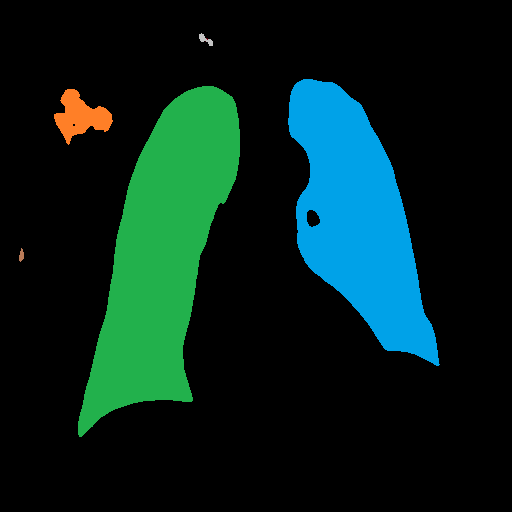}
\caption{An example of connected components of a lung image.}
\label{fig:connected_component}
\end{figure}

After heart and lung masks are computed, we performed dilation followed by erosion to fill holes in output mask \cite{soille1999}, then we find the connected components of prediction masks (Figure \ref{fig:connected_component}).

From the lung mask, we chose the two largest connected components and disregarded others as noise. The connected component with a lower x-axis coordinate is designated as the left lung mask and the other as the right lung mask. 
From the heart mask, we chose the connected component larger than a given threshold and closest to the center and designate it as the heart mask.

\subsection{Cardiothoracic Ratio Calculation}

\begin{figure*}[htbp]
\centering
\includegraphics[width=0.20\linewidth]{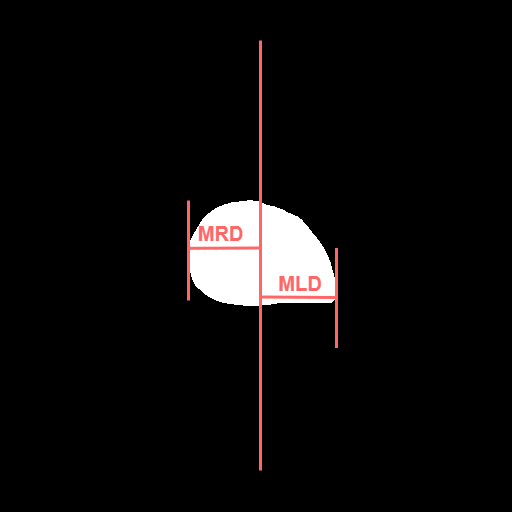}
\includegraphics[width=0.20\linewidth]{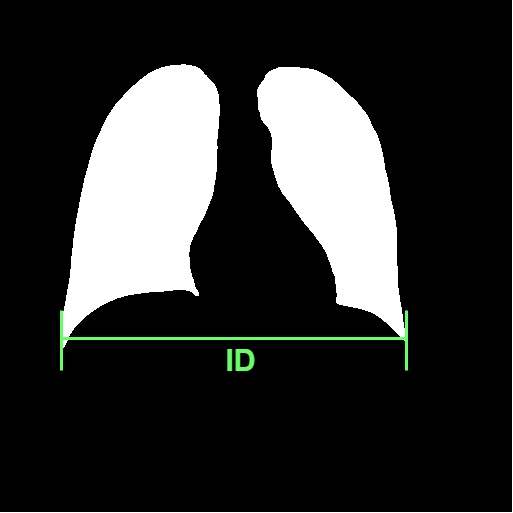}
\includegraphics[width=0.20\linewidth]{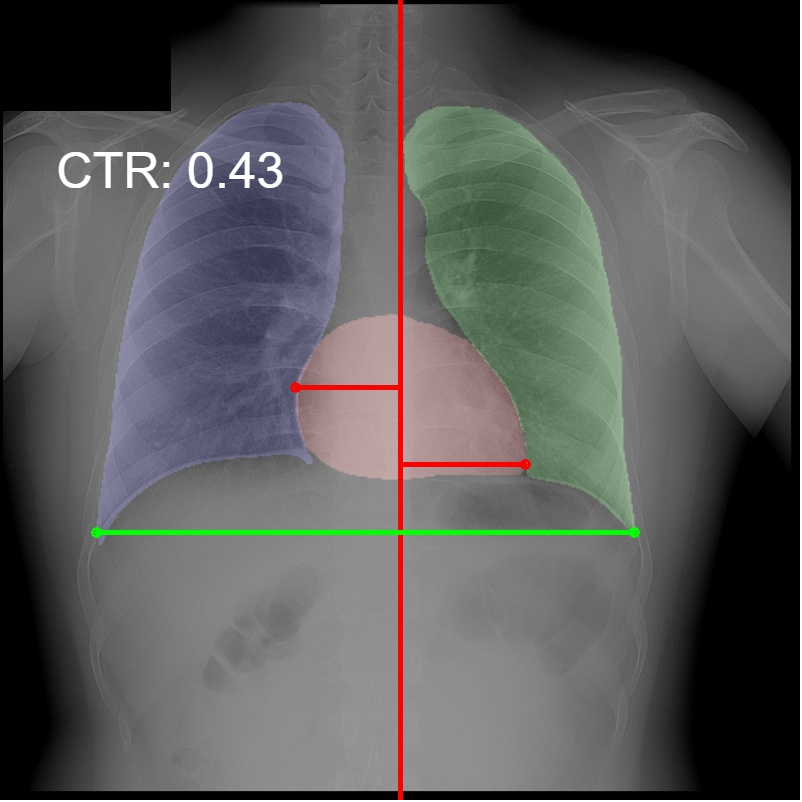}
\caption{CTR calculation by extents of heart and lung masks. $MRD$ and $MLD$ are calculated from heart mask, and $ID$ is calculated from lung mask.}
\label{fig:ctr_calculation}
\end{figure*}

After obtaining the masks, we calculate CTR with equation \ref{eqn:ctr}. We calculate cardiac diameter, $(MRD + MLD)$, by finding the extreme points on the x-axis of the heart mask and calculate their x-axis distance.
Thoracic diameter, $ID$, is calculated from extreme points from the lung mask.
Figure \ref{fig:ctr_calculation} demonstrates the calculation of each component.

\subsection{Cardiomegaly Detection}

In typical diagnostic practice, a normal measurement of CTR should be less than $0.5$, with CTR of $0.5$ to $0.55$ regarded as mild cardiomegaly and CTR of more than $0.55$ regarded as cardiomegaly \cite{Dimopoulos2008cardiomegaly}.
Since mild cardiomegaly is also mentioned in radiologist reports and is regarded as cardiomegaly, we use the ratio of $0.50$ as a cutoff value for the evaluation of cardiomegaly.

%% file: 4_results.tex
\section{Experiments and Results}




\subsection{Chest X-Ray Segmentation}
We compared our chest X-ray segmentation approach with Segnet \cite{badrinarayanan2017segnet} and U-Net \cite{olaf2015unet} without VGG16 encoder trained with the same samples and hyper-parameters.

Table \ref{table:seg_comparison} shows the result of our segmentation algorithms. Average Intersection-over-Union (IoU) values are calculated for each model on validation sets. Our U-Net+VGG16 model outperformed other approaches on heart segmentation, while producing comparable result to U-Net on lung segmentation.

Our U-Net+VGG16 lung segmentation model yields DSC (Dice's Coefficient) of 0.970, while the heart segmentation DSC was 0.892. Our lung and heart segmentation algorithm yields comparable results to studies performed on other data sets \cite{jangamsegmentation, candemir2013lung, wei2018scan}.

\begin{table}[htbp]
\begin{center}
\caption{Comparison of Intersection-over-Union (IoU) values of segmentation approaches}
\begin{tabular}{|c|c|c|}
\hline
\textbf{Model}&\textbf{Heart Segmentation}&\textbf{Lung Segmentation}\\
\hline
Segnet&$0.903$&$0.955$\\
\hline
U-Net&$0.907$&$0.965$\\
\hline
U-Net+VGG16&$0.919$&$0.963$\\
\hline
\end{tabular}
\label{table:seg_comparison}
\end{center}
\end{table}

\begin{figure*}[htbp]
\begin{center}
\begin{tabular}{c c c c c c}
\textbf{Lung Image}&\textbf{Ground Truth}&\textbf{Segnet}&\textbf{U-Net}&\textbf{U-Net+VGG16}\\
&&&&\\
\raisebox{-.5\totalheight}{\includegraphics[width=0.15\linewidth]{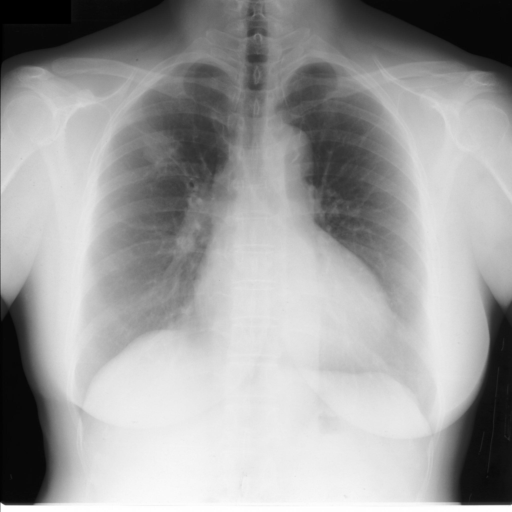}}
&\raisebox{-.5\totalheight}{\includegraphics[width=0.15\linewidth]{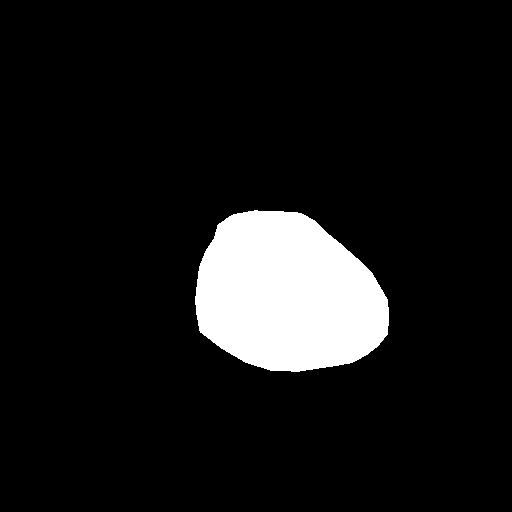}}
&\raisebox{-.5\totalheight}{\includegraphics[width=0.15\linewidth]{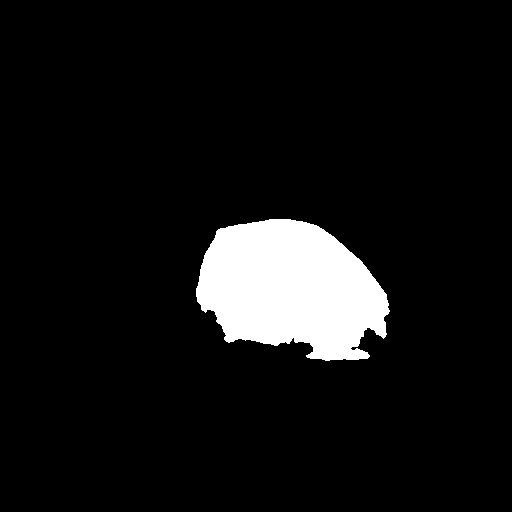}}
&\raisebox{-.5\totalheight}{\includegraphics[width=0.15\linewidth]{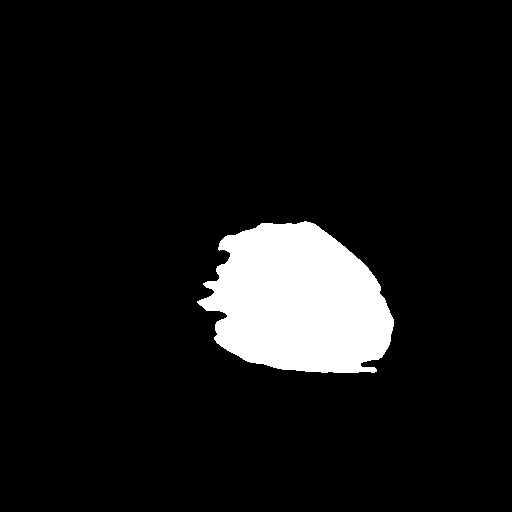}}
&\raisebox{-.5\totalheight}{\includegraphics[width=0.15\linewidth]{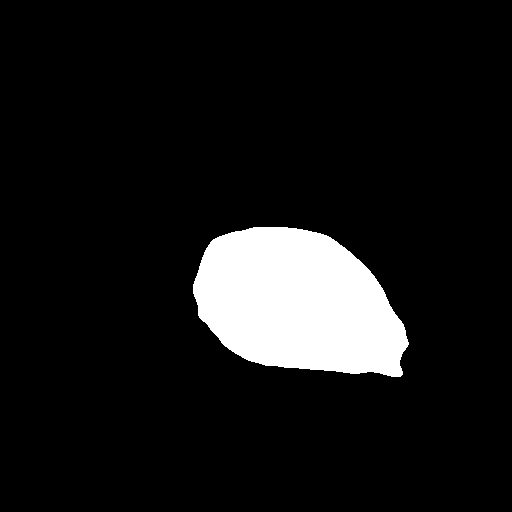}}\\[-1ex]
&&&&\\
\raisebox{-.5\totalheight}{\includegraphics[width=0.15\linewidth]{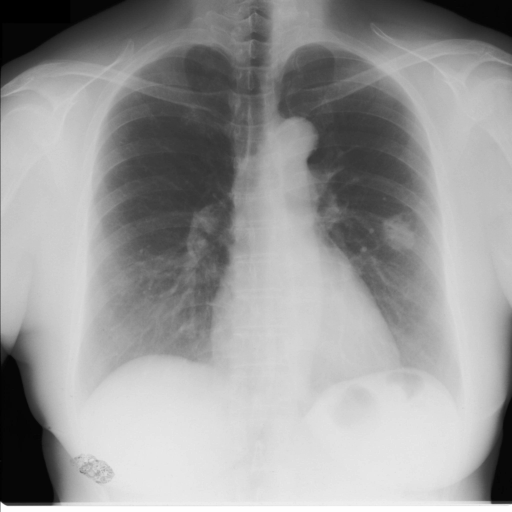}}
&\raisebox{-.5\totalheight}{\includegraphics[width=0.15\linewidth]{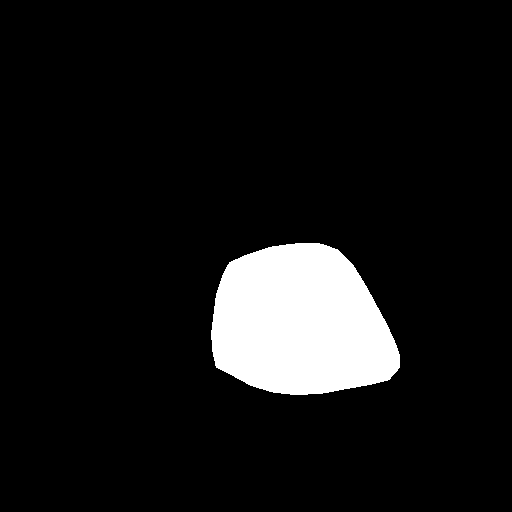}}
&\raisebox{-.5\totalheight}{\includegraphics[width=0.15\linewidth]{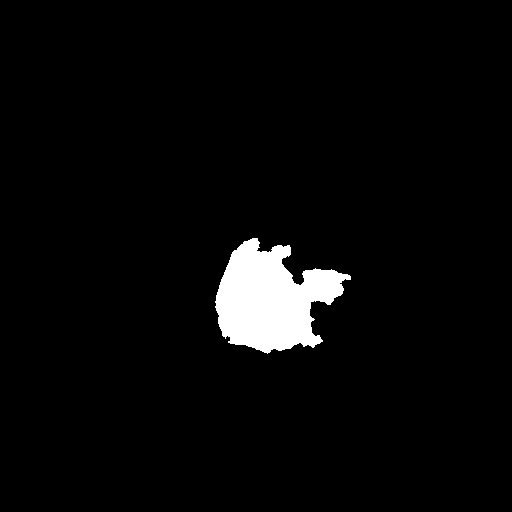}}
&\raisebox{-.5\totalheight}{\includegraphics[width=0.15\linewidth]{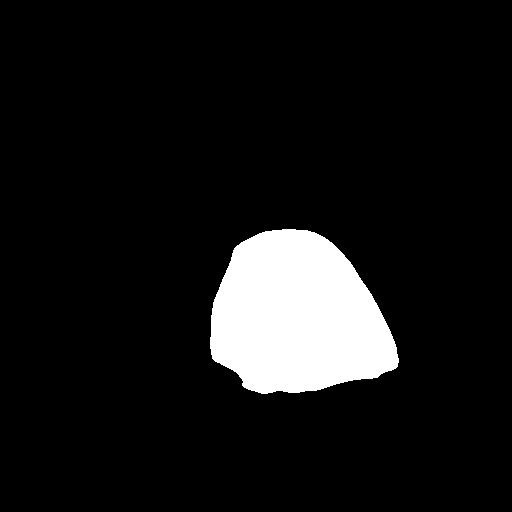}}
&\raisebox{-.5\totalheight}{\includegraphics[width=0.15\linewidth]{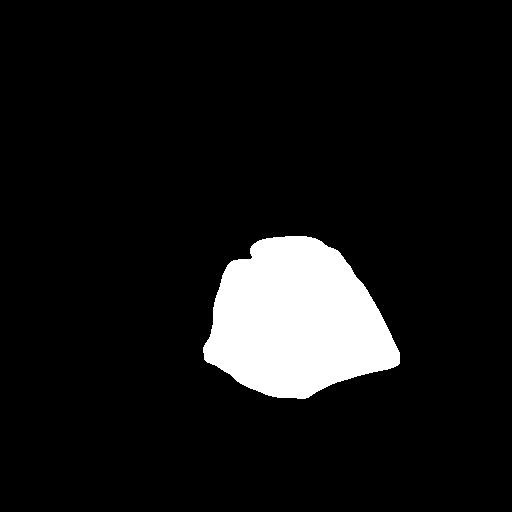}}\\[-1ex]
&&&&\\
\raisebox{-.5\totalheight}{\includegraphics[width=0.15\linewidth]{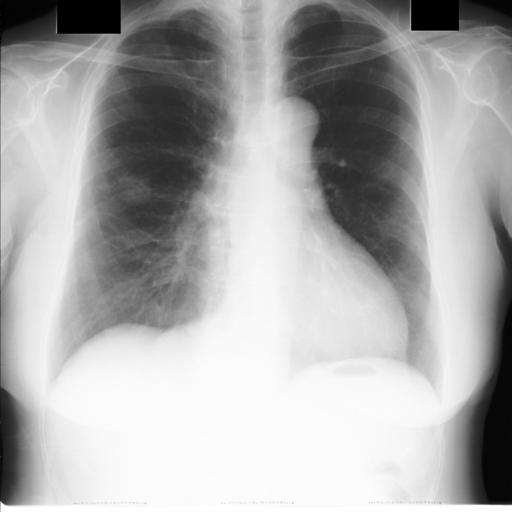}}
&\raisebox{-.5\totalheight}{\includegraphics[width=0.15\linewidth]{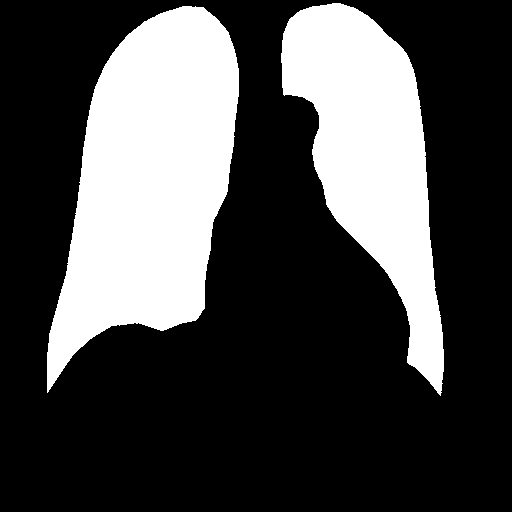}}
&\raisebox{-.5\totalheight}{\includegraphics[width=0.15\linewidth]{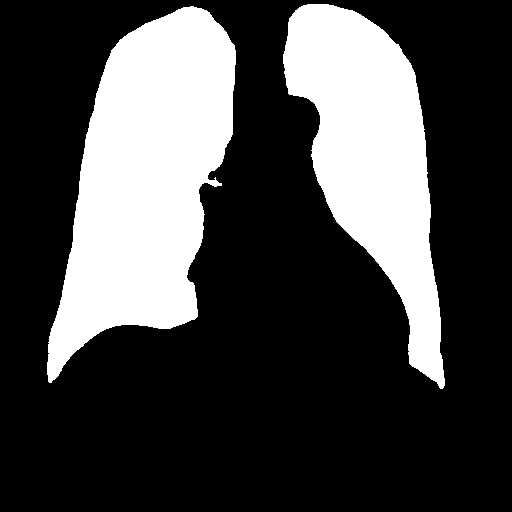}}
&\raisebox{-.5\totalheight}{\includegraphics[width=0.15\linewidth]{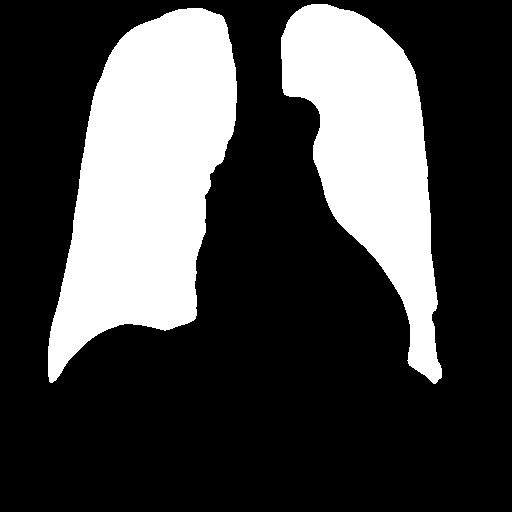}}
&\raisebox{-.5\totalheight}{\includegraphics[width=0.15\linewidth]{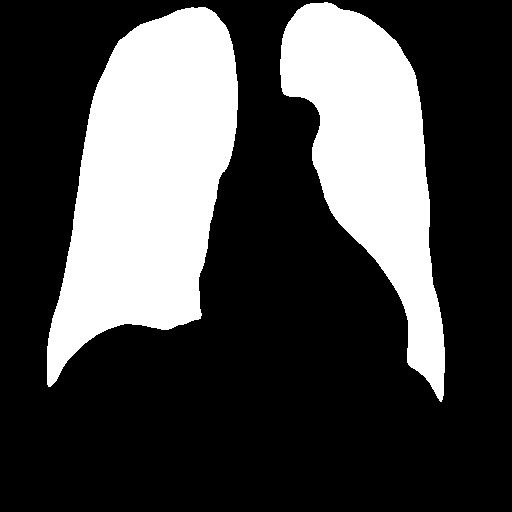}}\\[-1ex]
&&&&\\
\raisebox{-.5\totalheight}{\includegraphics[width=0.15\linewidth]{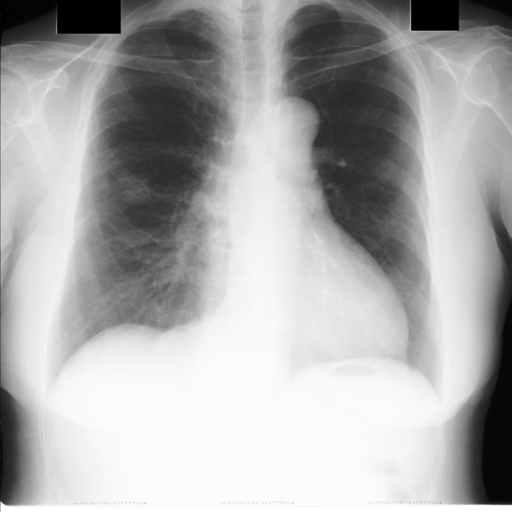}}
&\raisebox{-.5\totalheight}{\includegraphics[width=0.15\linewidth]{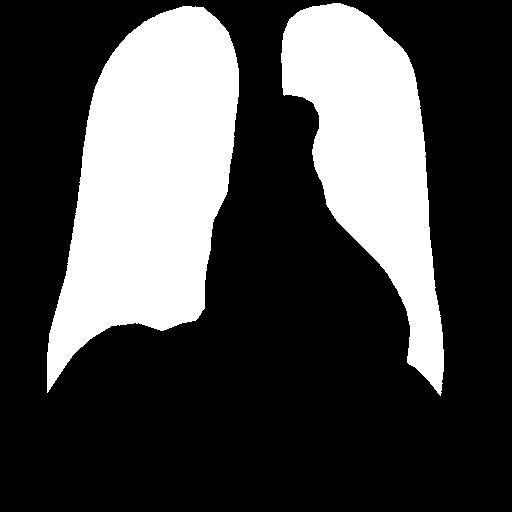}}
&\raisebox{-.5\totalheight}{\includegraphics[width=0.15\linewidth]{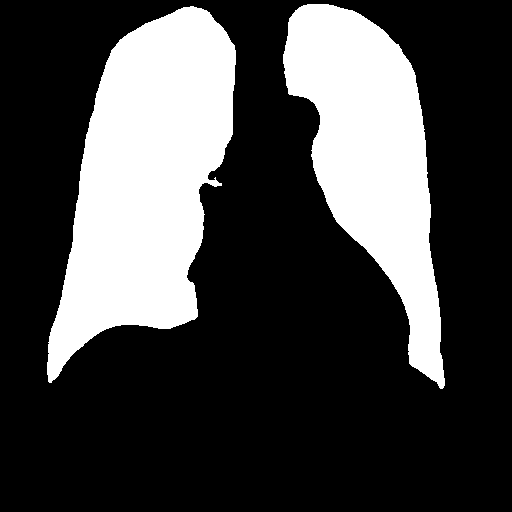}}
&\raisebox{-.5\totalheight}{\includegraphics[width=0.15\linewidth]{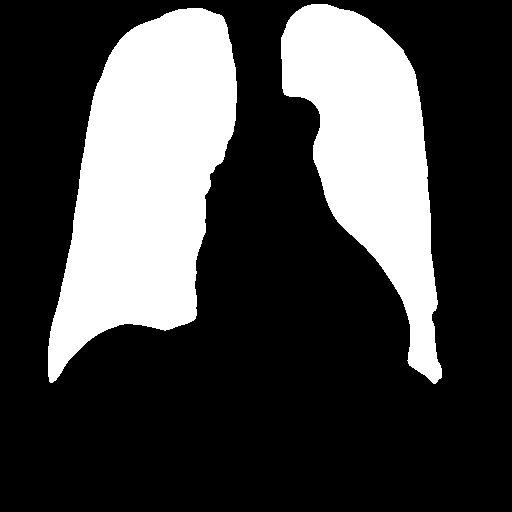}}
&\raisebox{-.5\totalheight}{\includegraphics[width=0.15\linewidth]{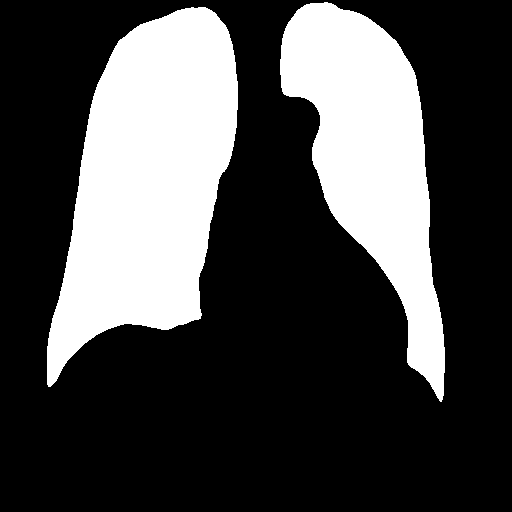}}\\[-1ex]
\end{tabular}
\caption{Example of heart and lung masks provided by Segnet, U-Net and U-Net with VGG16 Encoder models.}
\label{fig:seg_visual_comparison}
\end{center}
\end{figure*}

Figure \ref{fig:seg_visual_comparison} shows examples of heart and lung masks produced by each approach. U-Net with VGG16 encoder model provided the most consistent and smooth result.

\subsection{Cardiothoracic Ratio Calculation}

After obtaining heart and lung segments, we calculated $MRD$, $MLD$, and $ID$ and marked them on the image along with CTR value calculated from Equation \ref{eqn:ctr}.

\begin{figure}[htbp]
\centering
\includegraphics[width=0.4\linewidth]{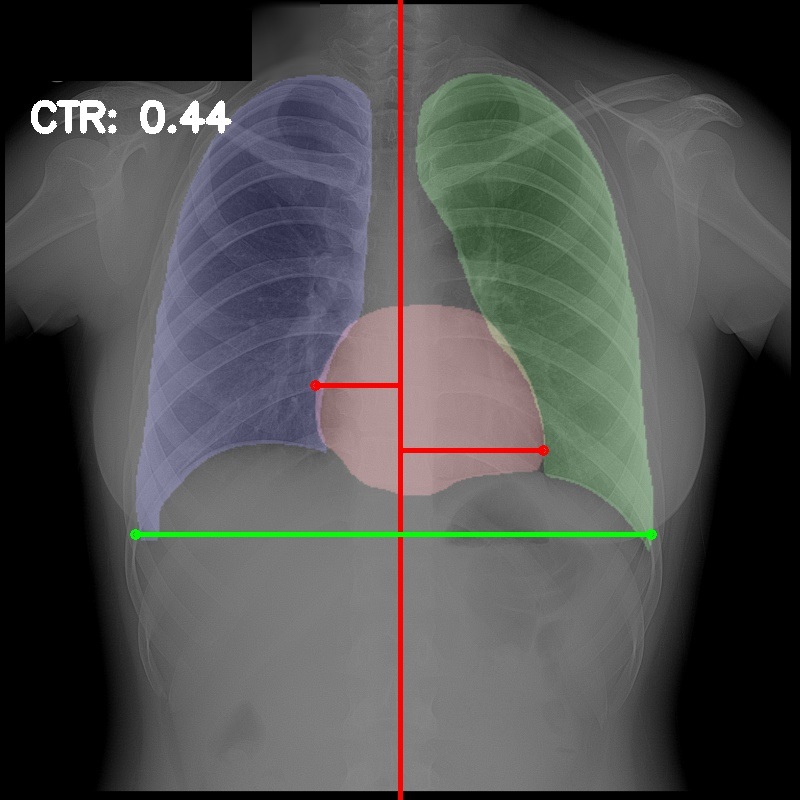}
\includegraphics[width=0.4\linewidth]{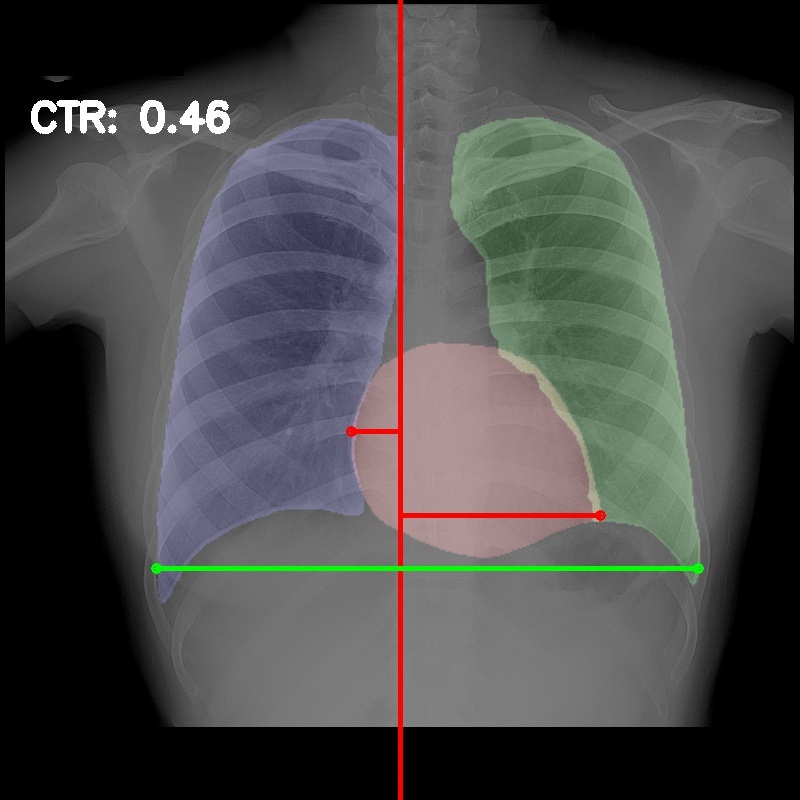}
\includegraphics[width=0.4\linewidth]{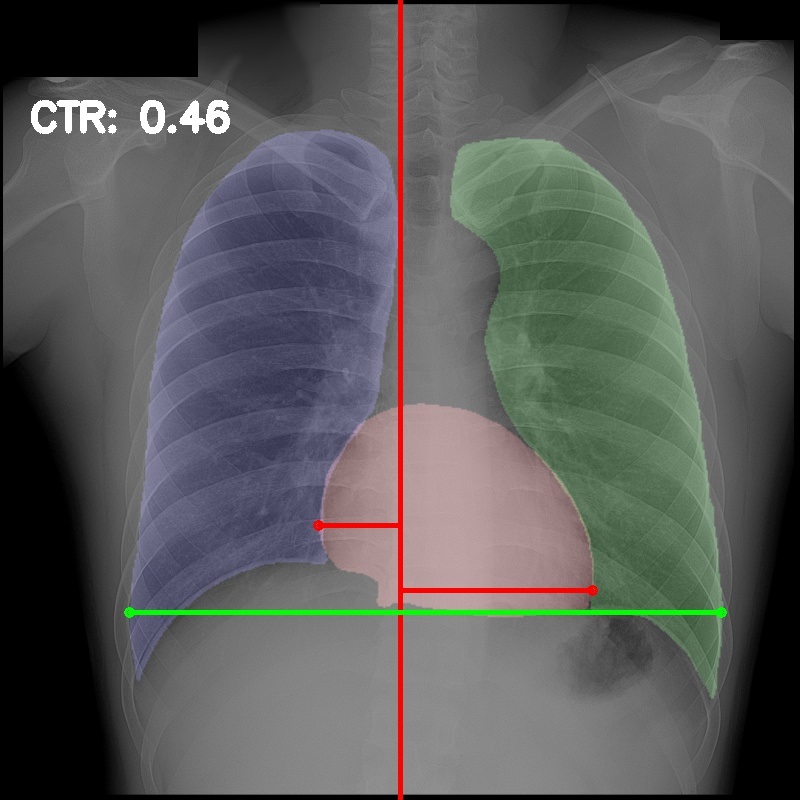}
\includegraphics[width=0.4\linewidth]{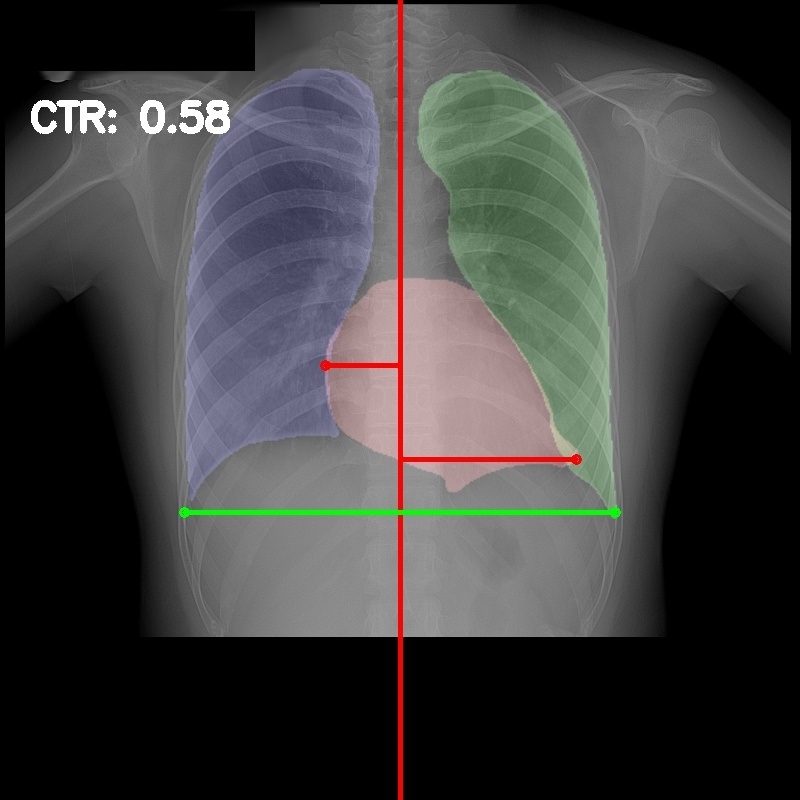}
\caption{Example of CTR calculation using our approach. Cardiomegaly condition is correctly identified in the bottom-right X-ray image.}
\label{fig:ctr_good_samples}
\end{figure}

Figure \ref{fig:ctr_good_samples} shows examples of our CTR calculation, where CTR values along with their $MRD$, $MLD$ and $ID$ measurement points are correctly identified. In this image, cardiomegaly is correctly detected by CTR value of $0.58$.

\begin{figure}[htbp]
\centering
\includegraphics[width=0.4\linewidth]{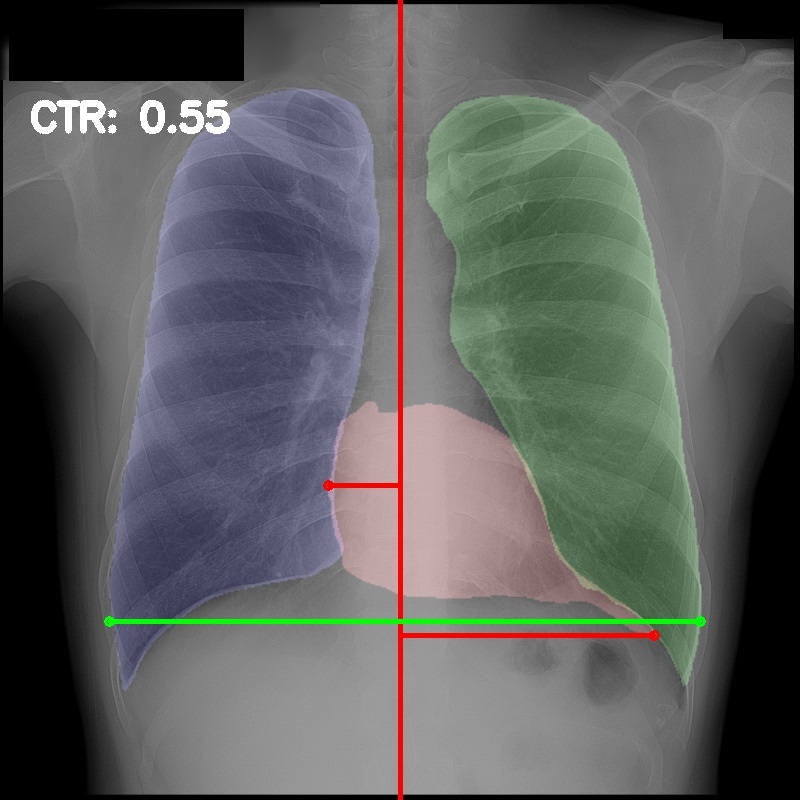}
\includegraphics[width=0.4\linewidth]{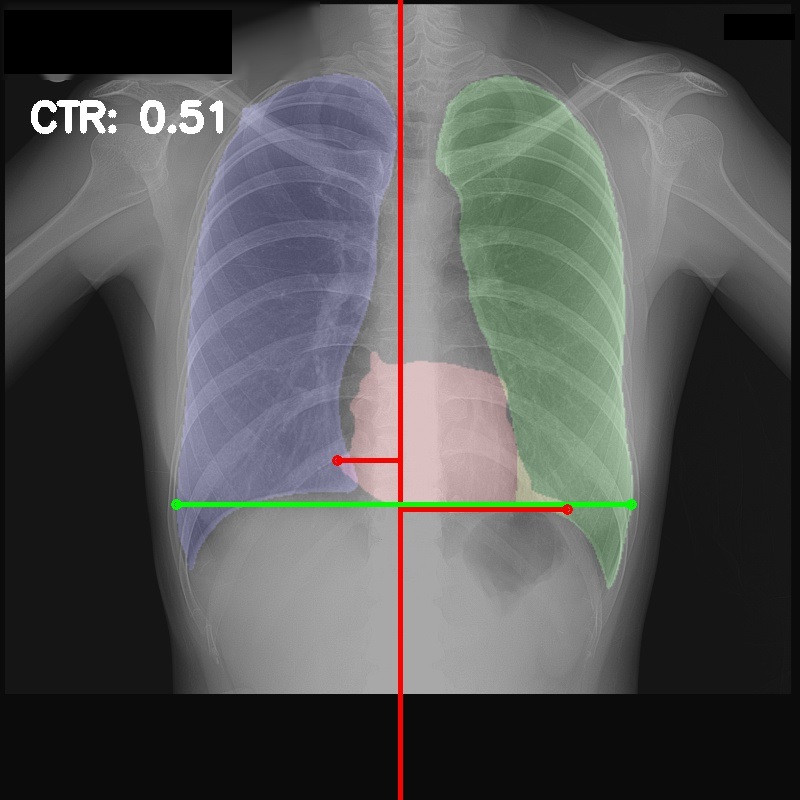}
\caption{Cases where our approach failed to properly calculate CTR. Heart mask regions are not properly detected and cause $MRD$ and $MLD$ values to be incorrectly computed.}
\label{fig:ctr_bad_samples}
\end{figure}

Figure \ref{fig:ctr_bad_samples} shows cases where our CTR calculation algorithm is incorrect due to failure to obtain accurate heart mask.

To measure the practical use of our CTR calculation, we asked human experts to verify if each of the measurements, $MRD$, $MLD$, and $ID$, are acceptable. We designate our result as correct if human experts indicate that our computer-generated measurements can be used in medical reports without any modification and designate them as incorrect otherwise.


We selected 600 images from NIH Chest X-ray and CheXpert datasets (total of 1,200 cases). For each dataset, we randomly sampled $300$ cases labeled with cardiomegaly positive and $300$ cases with no cardiomegaly. After filtering out images with technical issues and those in which human experts signify that the heart size cannot be accurately evaluated, we obtained a total of $1,022$ images, of which $491$ are with cardiomegaly label and $531$ are without cardiomegaly labels.

Table \ref{table:ctr_visual_comparison} shows the result of our experiment. Our approach can correctly measure $MRD$, $MLD$, and $ID$ in $76.5\%$ of the cases. There was no significant difference in algorithm's performance between cases with and without cardiomegaly conditions.

\begin{table}[htbp]
\begin{center}
\caption{Experiment on the practical use of our CTR calculation. Percent accuracy indicates the percentage of cases where human experts verify that all computer-generated measurements ($MRD$, $MLD$, and $ID$ values) are correct.}
\begin{tabular}{|c|c|c|c|}
\hline
\textbf{Category}&\textbf{Correct}&\textbf{Incorrect}&\textbf{Accuracy} \\
\hline
Cardiomegaly&$385$&$106$&$78.4\%$\\
\hline
No cardiomegaly&$397$&$134$&$74.8\%$\\
\hline
Total&$782$&$240$&$76.5\%$\\
\hline
\end{tabular}
\label{table:ctr_visual_comparison}
\end{center}
\end{table}

We also evaluated our CTR calculation approach in the degree to which its detection of cardiomegaly agrees with radiologist diagnosis of the condition. To this end, we selected $6004$ frontal AP view samples from NIH Chest X-ray dataset \cite{wang2017nihdataset} and the CheXpert dataset \cite{irvin2019chexpert} with specific mention of the presence or the absence of cardiomegaly. Cases where there is no mention of cardiomegaly are excluded from the analysis. In accordance to radiologists' practice, we predict that cardiomegaly is present if the CTR value is higher than the threshold of 0.5.  

We measured the detection accuracy, sensitivity, and specificity of our approach. We obtained the accuracy of $67.1\%$ and $69.8\%$ from the NIH Chest X-ray dataset and the CheXpert dataset, respectively. Full results are presented in Table \ref{table:ctr_comparison_acc}.

\begin{table}[htbp]
\begin{center}
\caption{Cardiomegaly detection performance using our automated CTR approach with a cut-off at 0.5}
\begin{tabular}{|c|c|c|c|}
\hline
\textbf{Dataset}&\textbf{Accuracy}&\textbf{Sensitivity}&\textbf{Specificity} \\
\hline
NIH Chest X-ray&$67.1\%$&$0.81$&$0.69$\\
\hline
CheXpert&$69.8\%$&$0.69$&$0.70$\\
\hline
\end{tabular}
\label{table:ctr_comparison_acc}
\end{center}
\end{table}


In practice, CTR values near threshold ($0.5$) are prone to produce diagnostic errors. Therefore we analyzed the distribution of images at different ranges of CTR values to confirm this hypothesis. Table \ref{table:ctr_comparison_range} shows the distributions from NIH Chest X-Ray and CheXpert dataset. $83.3\%$ and $87.4\%$ of false positive samples in NIH Chest X-ray and CheXpert datasets, respectively, have CTR values between $0.5$ to $0.6$, suggesting that machine-generated labels in this range need to be re-evaluated by human readers. 

\begin{figure}[htbp]
\centering
\includegraphics[width=0.32\linewidth]{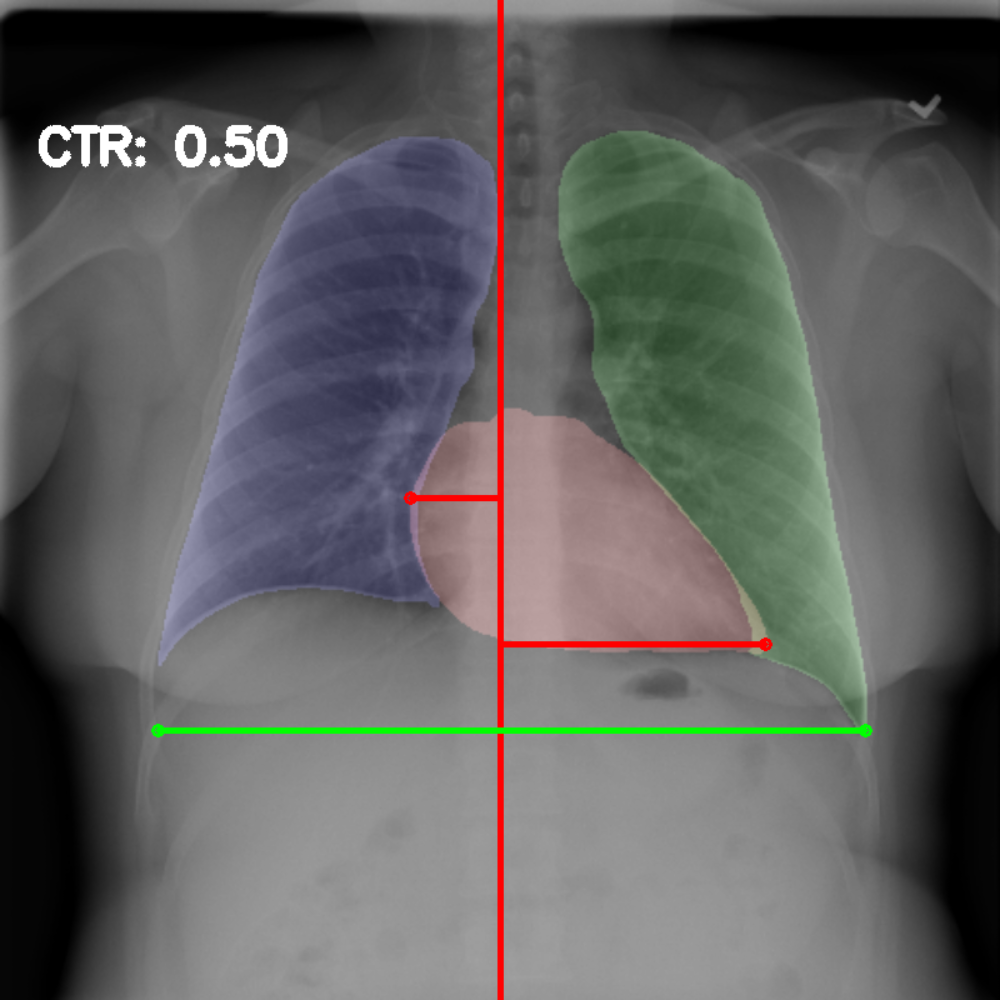}
\includegraphics[width=0.32\linewidth]{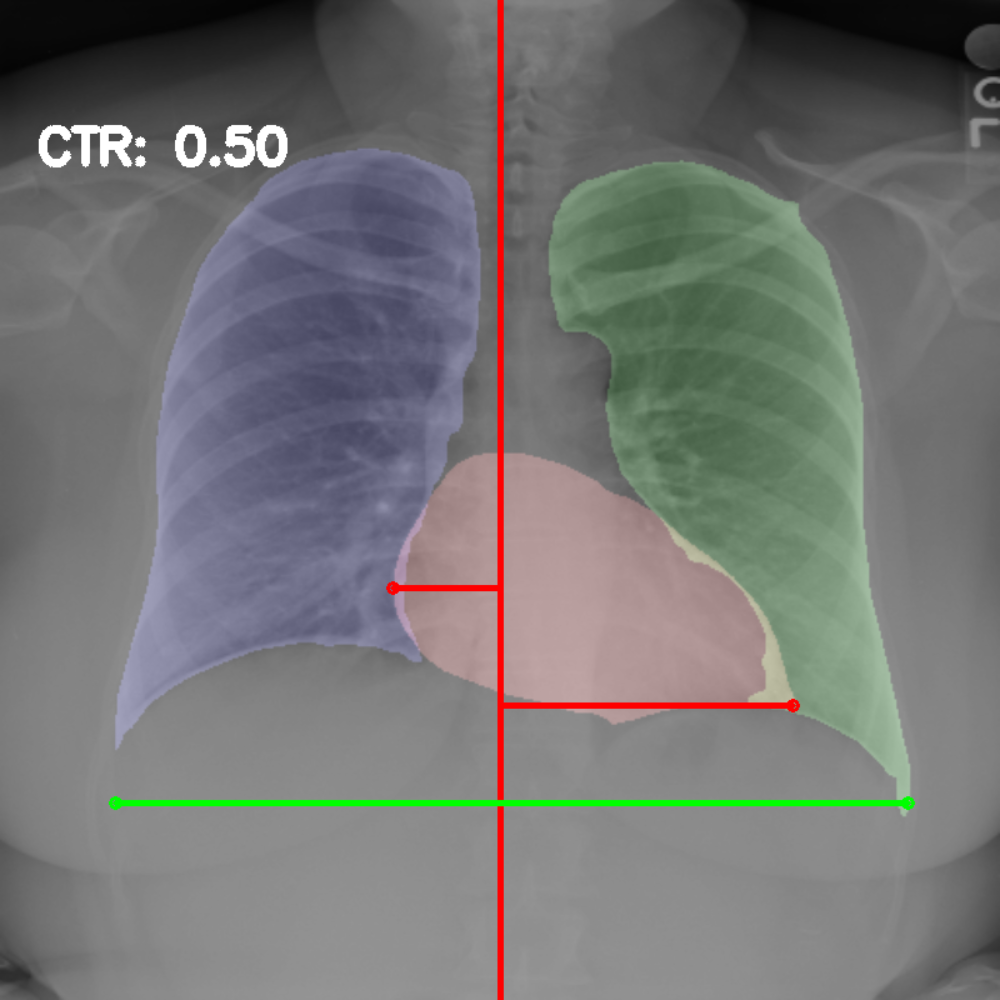}
\includegraphics[width=0.32\linewidth]{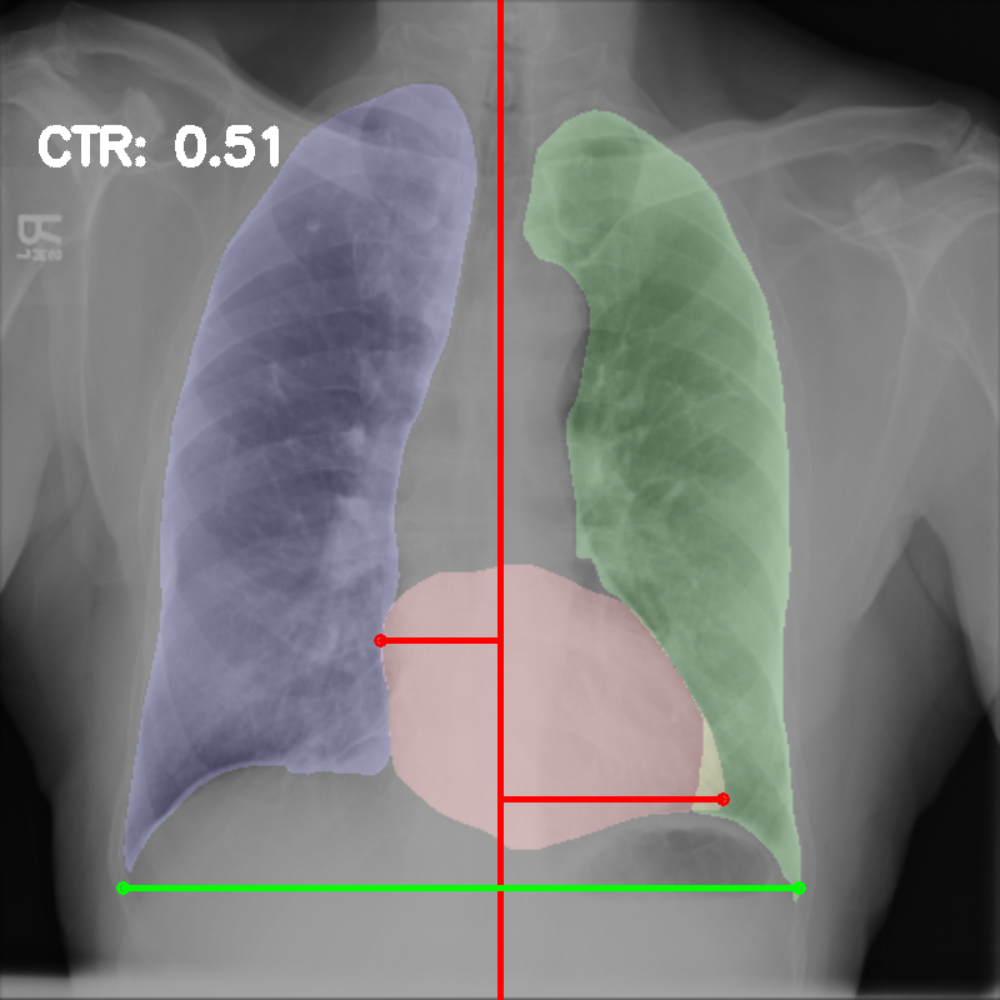}
\caption{Sample CXR images from the NIH Chest X-ray dataset containing no-cardiomegaly labels with detected CTR values on $0.5-0.55$ range.}
\label{fig:ctr_borderline}
\end{figure}

However, upon manual inspection of the images, we found a relatively high number of mild cardiomegaly cases that are not included in the dataset labels. Figure \ref{fig:ctr_borderline} shows sample chest X-ray images containing no-cardiomegaly labels with detected CTR values on $0.5-0.55$ range. This reflects the general sentiment that dataset labels are noisy and may contain errors, making it harder to analyze our algorithm. It also illustrates the value of an algorithmic "second opinion" to catch potential errors from both human and NLP algorithms used to generate the labels.


\begin{table}[htbp]
\begin{center}
\caption{Distribution of images at different ranges of CTR values. The CTR values are those given by our algorithm while Pos and Neg labels are those provided by dataset.}
\begin{tabular}{|p{2.4cm}|p{0.5cm}|p{1.1cm}|p{1.8cm}|p{1.8cm}|p{1.8cm}|p{1.8cm}|p{1.1cm}|}
\hline
\multicolumn{2}{|c|}{\multirow{2}{*}{\textbf{Dataset}}}&\multicolumn{6}{|c|}{\textbf{Distribution $(\%)$ at Ranges of CTR Values}}\\
\cline{3-8}
\multicolumn{2}{|c|}{}&$<0.40$&$0.40-0.45$&$0.45-0.50$&$0.50-0.55$&$0.55-0.60$&$>0.60$\\
\hline
\multirow{2}{*}{NIH Chest X-ray}&Pos&$2.8$&$2.1$&$9.4$&$30.1$&$29.3$&$26.2$\\
\cline{3-8}
&Neg&$15.1$&$24.3$&$27.5$&$18.6$&$8.9$&$5.5$\\
\hline
\multirow{2}{*}{CheXpert}&Pos&$14.3$&$5.9$&$9.0$&$15.5$&$22.1$&$33.3$\\
\cline{3-8}
&Neg&$12.6$&$16.5$&$26.4$&$25.1$&$13.8$&$5.6$\\
\hline
\end{tabular}
\label{table:ctr_comparison_range}
\end{center}
\end{table}


\begin{table}[htbp]
\begin{center}
\caption{Analysis to demonstrate the mismatch between human expert and dataset labels. Analysis was conducted on the subset of images where radiologists accept machine-generated CTR values. Percentages showed the portion of images where $CTR<0.5$ with cardiomegaly labels and images where $CTR>=0.5$ with no-cardiomegaly labels}
\begin{tabular}{|c|c|c|c|c|}
\hline
\textbf{Annotation}&\textbf{Dataset}&\textbf{CTR$<$0.5}&\textbf{CTR$>=$0.5}&\textbf{Errors}\\
\hline
\multirow{3}{2.5cm}{NIH Chest X-ray}
&Cardiomegaly&$19$&$194$&$8.9\%$\\
&No Cardiomegaly&$172$&$40$&$18.9\%$\\
\cline{2-5}
&Average&&&$13.9\%$\\
\hline
\multirow{3}{1.9cm}{CheXpert}
&Cardiomegaly&$42$&$130$&$24.4\%$\\
&No Cardiomegaly&$110$&$75$&$40.5\%$\\
\cline{2-5}
&Average&&&$32.5\%$\\
\hline
\end{tabular}
\label{table:dataset_validation}
\end{center}
\end{table}

To demonstrate the mismatch between human expert and dataset labels, we performed analysis on $782$ samples where human experts accepted our $MRD$, $MLD$, and $ID$ measurements. Table \ref{table:dataset_validation} shows the result of this analysis. It can be seen that $13.9\%$ and $32.5\%$ of the labels from NIH Chest X-ray and CheXpert dataset, respectively, did not agree with CTR values accepted by local experts.

The experiment also shows that there are more mismatches from cases with no-cardiomegaly labels, with $18.9\%$ and $40.5\%$ mismatches from NIH Chest X-ray dataset and CheXpert dataset, respectively, compared to cases with cardiomegaly labels, with $8.9\%$ and $24.4\%$ respectively. This suggested that computer algorithms can pick up mild cardiomegaly cases that were not indicated by dataset labels.

%% file: 5_conclusion.tex
\section*{Conclusion}
Our work presents a simple approach to evaluate CTR automatically from chest X-ray images. In this preliminary research, we were able to achieve $76.5\%$ acceptance rate in practical settings, which translates to the amount of the time saved for radiologists from measuring cardio and thoracic diameters and calculating CTR manually. Apart from saving a significant amount of time for radiologists, our approach has additional benefits in that it can alert radiologists on cases with cardiomegaly that are not obvious to the human eyes and thus provide an algorithmic "second opinion". Our simple approach can be integrated into a CTR assessment tool, which can display clear positions where $MRD$, $MLD$, and $ID$ measurements are made, so that radiologists can confirm the results immediately. Our deep learning algorithm is able to achieve high accuracy on images obtained from different hospitals even when trained with less than $400$ labeled samples. With more samples, we are confident that the accuracy will improve.